\documentclass[a4paper]{jpconf}
\usepackage{graphicx}

\newcommand{\mtop}{\ensuremath{m_{\mathrm{top}}}}
\newcommand{\mlb}{\ensuremath{m(\ell b)}}
\def\antibar#1{\ensuremath{#1\bar{#1}}}

\def\ttbar{\antibar{t}}

\def\TeV{\ifmmode {\mathrm{\ Te\kern -0.1em V}}\else
                   \textrm{Te\kern -0.1em V}\fi}%
\def\GeV{\ifmmode {\mathrm{\ Ge\kern -0.1em V}}\else
                   \textrm{Ge\kern -0.1em V}\fi}%
\def\MeV{\ifmmode {\mathrm{\ Me\kern -0.1em V}}\else
                   \textrm{Me\kern -0.1em V}\fi}%

\def\pt{\ensuremath{p_{\mathrm{T}}}} % Subscript roman not italic (EE)
\def\pT{\ensuremath{p_{\mathrm{T}}}} % Subscript roman not italic (EE)
\def\MET{\ensuremath{E_{\mathrm{T}}^{\mathrm{miss}}}} % Sub/superscript roman not italic (EE)
\begin{document}
\title{\boldmath Measurement of the top quark mass in topologies enhanced with single top quarks produced in the $t$-channel at $\sqrt{s}=8\,\mathrm{TeV}$ using the ATLAS experiment}

\author{Hendrik Esch$^1$ On behalf of the ATLAS Collaboration.}

\address{$^1$ TU Dortmund, Exp. Physik IV, Otto-Hahn-Str. 4, 44227 Dortmund, Germany}

\ead{hendrik.esch@uni-dortmund.de}

\begin{abstract}
This article presents a measurement of the top quark mass in topologies enhanced with single top quarks produced in 
the $t$-channel produced via weak interactions. The dataset was collected at a centre-of-mass 
energy of $\sqrt{s}=8\,\mathrm{TeV}$ with the ATLAS detector at the LHC and corresponds to an integrated luminosity 
of $20.3\,\mathrm{fb^{-1}}$. 
To determine the top quark mass a template method is used based on the distribution of the invariant mass of
the lepton and the $b$-tagged jet as estimator. 
The result of the measurement is $m_{\mathrm{top}} = 172.2 \pm 0.7 {\mathrm{(stat.)}} \pm 2.0 {\mathrm{(syst.)}}\,\mathrm{GeV}$.
\end{abstract}

\section{Introduction}
\label{sec:intro}

In addition to the \ttbar{} pair production via the strong interaction, in proton-proton ($pp$) collisions at the LHC,
top quarks can also be produced singly via the weak charged-current interactions, giving another possibility for measuring the top quark mass.
The dominant process for single top quark production is the $t$-channel exchange of a virtual $W$-boson. 
Important differences from the production mode compared to \ttbar{}, resulting in different sizes of certain systematic uncertainties, and the fact that the
measurement of the top quark mass is obtained from a statistically independent sample, 
provide an excellent motivation for such a measurement and for including it in future combinations with other measurements.

In this article, the first measurement of $\mtop$ in topologies enhanced with $t$-channel single top quark production with the ATLAS experiment~\cite{ATLAS} is presented.
Production of top quark pairs also give a significant contribution to the sample, while $Wt$ production and $s$-channel production only give minor contributions. 
Events are characterised by an isolated high-$p_T$ charged lepton (electron or muon), 
missing transverse momentum from the neutrino and exactly two jets produced by the hadronisation of the $b$-quark and the light quark in the $t$-channel. 
The main backgrounds are $W/Z$+jets production, especially in association with heavy quarks, diboson production, and multijet production via QCD processes. 
Events from all single top production processes and \ttbar{} production are treated as signal in the analysis.

\section{Event selection}
\label{sec:objsel}

Based on the expected signature of the signal, events are selected with exactly one isolated electron or muon, 
missing transverse momentum and exactly two jets, out of which one is required to be identified as a $b$-quark jet.

Off\/line electron and muon candidates are required to be isolated and satisfy $\pT > 25\GeV$ and $|\eta| < 2.5$. 

Jets are reconstructed using the anti-$k_{t}$ algorithm with a distance parameter of 0.4. The reconstruction is based on locally 
calibrated clusters with simulation-based as well as in-situ calibrations based on data. 
They are required to satisfy $\pt > 30\GeV$ and $|\eta| < 4.5$. Jets within $2.75<|\eta|<3.5$, which have 
significant energy deposited in the endcap-forward calorimeter transition region, must have $\pT >35\GeV$.
Exactly one of the selected jets is required to be identified as a $b$-quark jet by the MV1c $b$-tagging algorithm. The algorithm is applied at an 
efficiency of 50\% for $b$-jets in simulated $\ttbar$ events.

In order to reduce the number of multijet background events, which are characterised by low $\MET$ and 
low transverse $W$-boson mass $m_{\mathrm{T}}(W)$, the event selection requires $\MET > 30\GeV$ and $m_{\mathrm{T}}(W) > 50\GeV$.
Another class of multijet background events are further reduced by applying an additional cut, which is realised by the following
condition between the lepton $p_{T}$ and the $\Delta \phi \left(j_1, \ell \right)$:
\begin{equation}
\pT\left(\ell\right) > 40\GeV \left(1 -  \frac{\pi - \left|\Delta \phi\left(j_1, \ell \right)\right|}{\pi -1} \right),
\end{equation}
where $\ell$ denotes the identified charged lepton and $j_1$ the reconstructed jet with the highest $\pT$~\cite{ATLAS-CONF-2014-055}.

\section{Background Estimation}
\label{sec:background}

To determine the normalisation of the multijets background, a binned maximum likelihood fit is performed to the \MET{} distribution in data 
after applying all selection criteria, with the cut on \MET{} removed. Template distributions for the multijet backgroud are obtained by different methods in 
the electron and muon channel, correspondingly.

In the electron channel a jet-lepton model~\cite{ATLAS-CONF-2014-058} is obtained by selecting 
simulated multijet events with jets that have similar properties to selected electrons.
In the muon channel an anti-muon method~\cite{ATLAS-CONF-2014-058} is used, which builds a multijet model derived from collision data.

The multijet template is fitted together with templates derived from MC simulation for all other processes
whose rate uncertainties are accounted for in the fitting process in the form of additional constrained nuisance parameters. 
The corresponding \MET{} distributions after rescaling the different backgrounds and the multijet template 
to their respective fit results are shown in~\cite{ATLAS-CONF-2014-007} for both the electron and muon channel. 

\section{Neural network selection}
\label{sec:nn}

Following the event selection described in Section~\ref{sec:objsel}, the selected sample is still dominated by background processes.
Multivariate analysis techniques are used to separate signal from background candidates.
A neural network classifier \cite{Feindt:2006pm} that combines a three-layer feed-forward neural network with a preprocessing of the input variables 
is used to enhance the separation power.

The network infrastructure consists of one input node for each input variables plus one bias node, 15 nodes in the hidden layer,
and one output node which gives a continuous output in the interval [$0,1$].
The training is done using single top $t$-channel events as signal during the training and $W+$jets,
$Z+$jets, and diboson processes are considered as background. Extensive studies were done to ensure
that using a signal sample with a fixed top quark mass does not bias the result of the measurement.

The input variables to the neural network are selected so that for a minimal number of 
variables the best possible separation between the signal and background processes is achieved.
Each variable is initially tested for agreement between the MC background model and observed data events in 
control regions and, taking into account potential signal contributions, is also tested in the signal region. 
\vspace*{-0.25cm}

\begin{figure}[h]
\begin{minipage}{17pc}
\includegraphics[width=17pc]{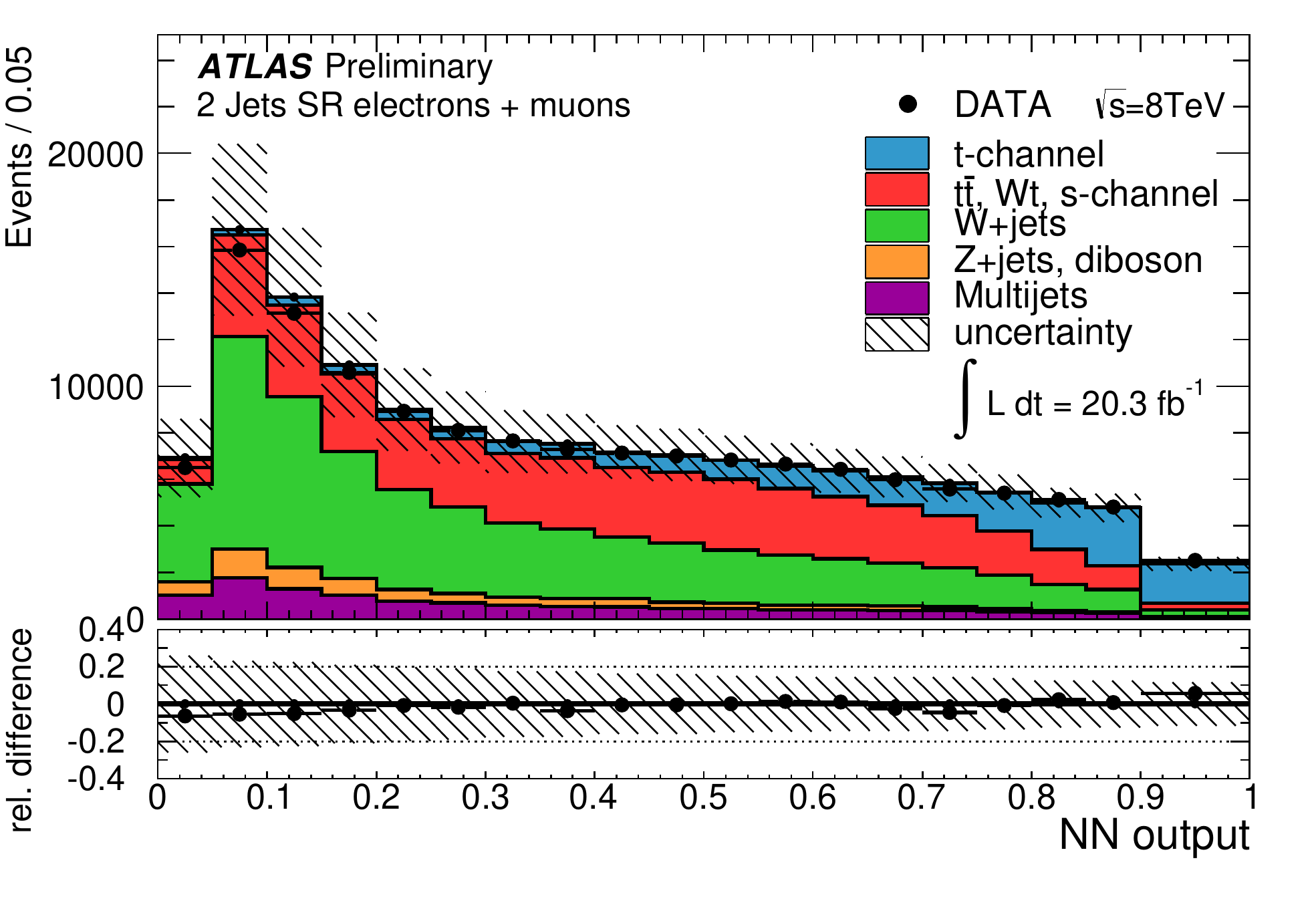}
\caption{\label{fig:finalnn} Neural network output distribution in the signal region normalised to the number of expected events~\cite{ATLAS-CONF-2014-055}.}
\end{minipage}\hspace{2pc}%
\begin{minipage}{19pc}
This leads to 12 variables remaining for the network including variables obtained from the reconstructed $W$-boson and the top quark. 

\hspace{1.5em}The resulting neural network output distributions for the various processes in the signal region
is shown in Figure~\ref{fig:finalnn}. Signal-like events have output values close to one, whereas background-like events accumulate near zero. 

\hspace{1.5em}To enhance the signal sample with single top and \ttbar{} events a cut on the neural 
network output variable at $0.75$ is chosen. In the signal region 19833 events that fulfill this cut are observed 
in data while the expectation from SM backgrounds amounts to $19470 \pm 2700$ events.
\end{minipage} 
\end{figure}
\vspace*{-0.5cm}

\section{\boldmath Measurement of \mtop{} with a template method}
\label{sec:template}

In order to measure the top quark mass in the signal region after the cut on the neural network, a template method is used. Simulated
distributions are constructed for \mlb{}, which is sensitive to the top quark mass, using a number of discrete values of \mtop{}. 
This \mlb{} estimator is defined as the invariant mass of the charged lepton plus the $b$-jet system.

\begin{figure}[!htbp]
\begin{minipage}{17pc}
\includegraphics[width=17pc]{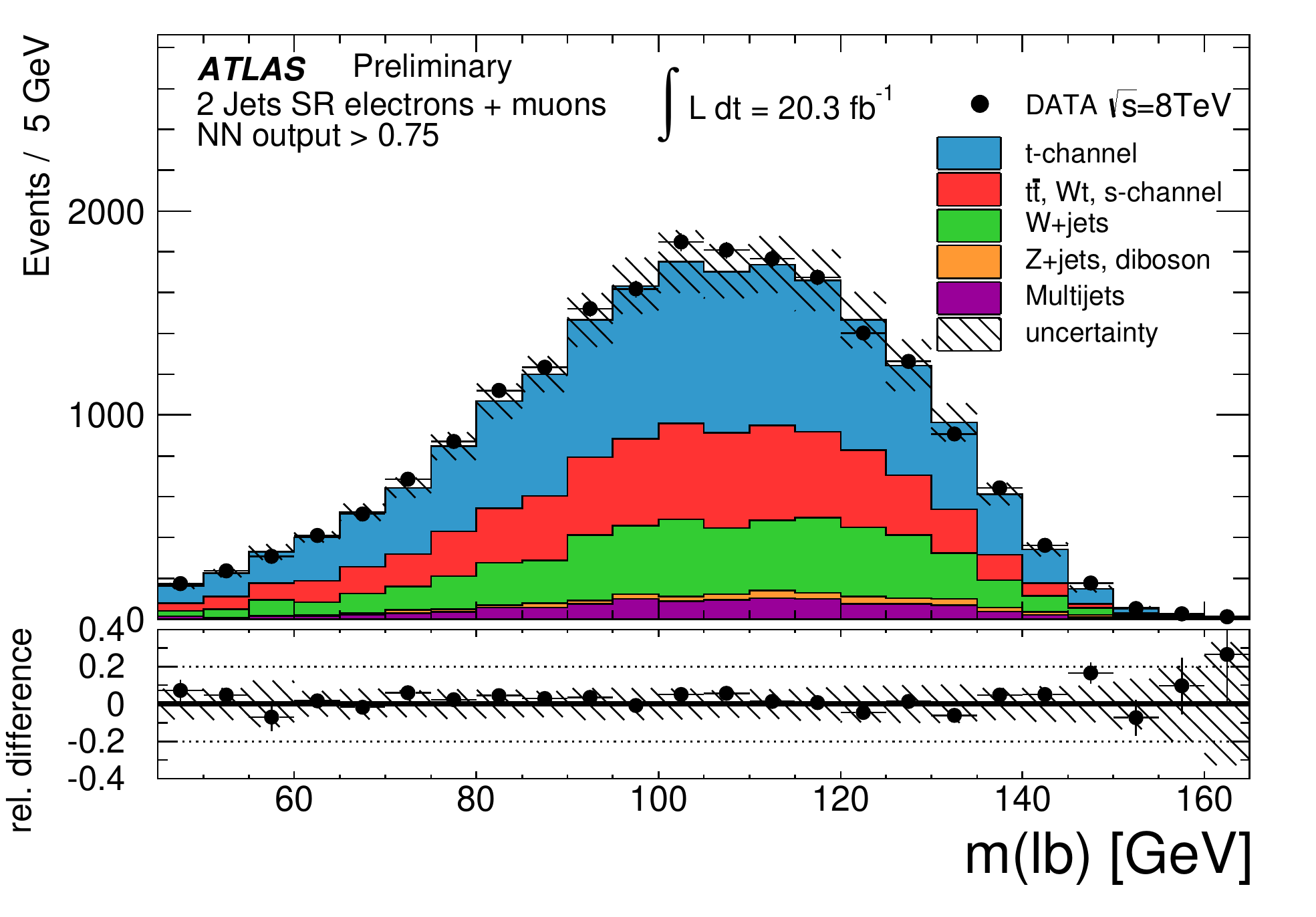}
\caption{\label{fig05a}Distributions of $\mlb$ for events with an output value of the neural network larger than $0.75$. The signal MC processes assume $\mtop = 172.5\GeV$ and the expected 
distribution is normalised to the number of expected events~\cite{ATLAS-CONF-2014-055}.}
\end{minipage}\hspace{2pc}%
\begin{minipage}{17pc}
\includegraphics[width=17pc]{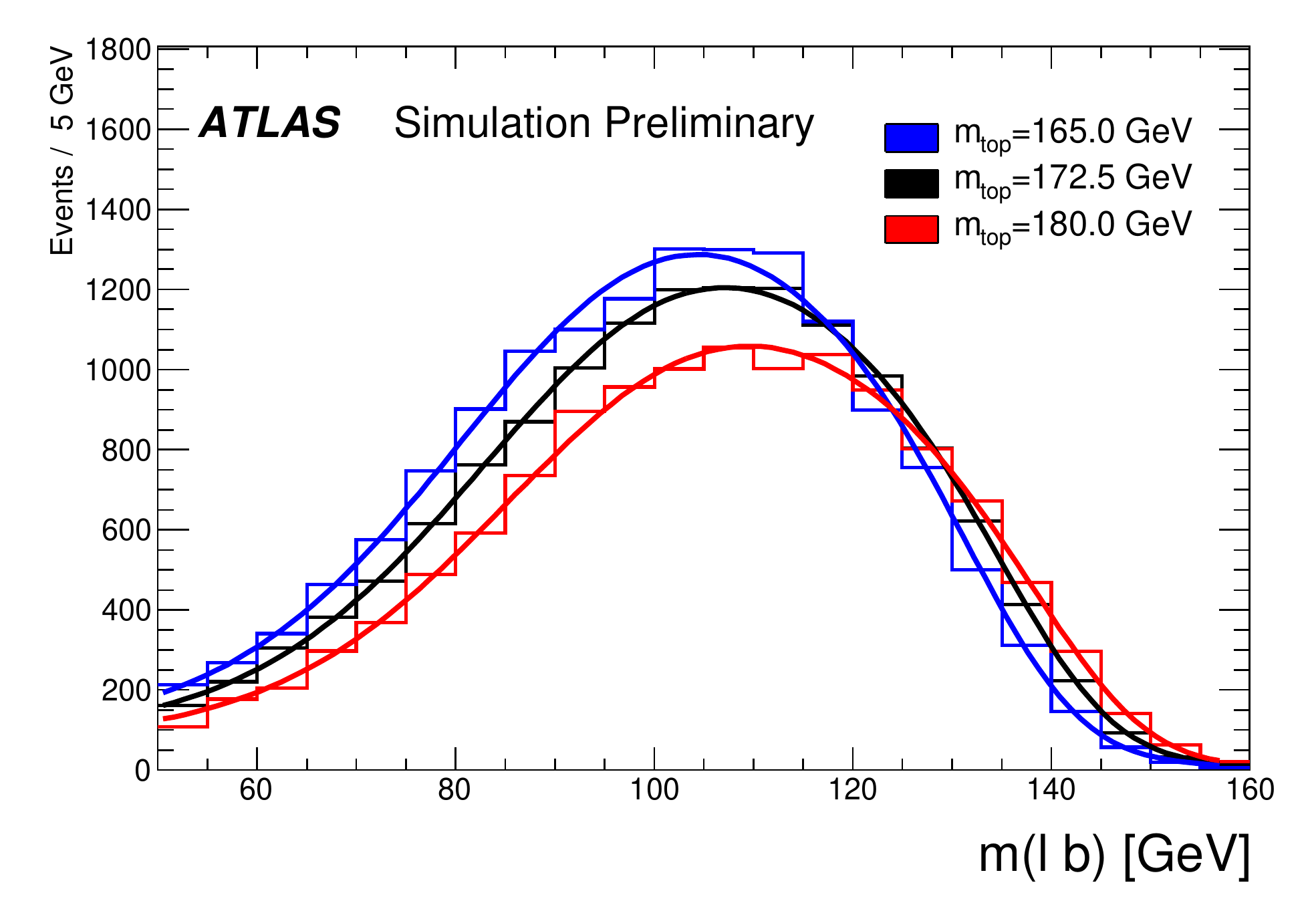}
\caption{\label{fig05b}Dependence of the $\mlb$ distribution of all top quark processes on mtop for the signal MC samples generated with different input top quark masses, together 
with the signal probability density functions~\cite{ATLAS-CONF-2014-055}.}
\end{minipage} 
\end{figure}

The resulting distribution in the signal region after the cut on the neural network output in data together with the prediction assuming $\mtop = 172.5\,\GeV$ is shown in Figure~\ref{fig05a}. 

The templates are parametrised and the parameters are then interpolated between different values of \mtop{}. In Figure~\ref{fig05b} the sensitivity of the \mlb{} observable to the input 
value of the top quark mass is shown by the \mlb{} distributions for three different mass points together with their respective fitted parametrisations. All single-top and \ttbar{} processes are 
treated as signal and the signal templates for \mlb{} are fitted using an analytic expression including Landau and Gauss parametrisations.
The same parametrisation is used for the mass-independent \mlb{} distribution of the background, which is dominated by W+jets and QCD-multijet production.

In the final step a likelihood fit to the observed data distribution is used to obtain the value of \mtop{} that best describes the data. 
The likelihood has three parameters: the top quark mass \mtop{}, the relative background fraction $f$, and the overall normalisation $N$. 
The fraction $f$ is constrained by a Gaussian distribution centred around the prediction from simulation $f_{\mathrm{bkg}}$. The width of the Gaussian $\sigma_{f_{\mathrm{bkg}}}$ 
reflects the theoretical uncertainty on the background fraction. 

\section{Results}
\label{sec:result}

The result of the fit to 2012 ATLAS data in topologies enhanced with $t$-channel single top quarks events is:
\begin{equation}
 \mtop = 172.2 \pm 0.7\,(\mathrm{stat.}) \pm 2.0\,(\mathrm{syst.})\GeV.
\end{equation}

\begin{figure}[h]
\includegraphics[width=16pc]{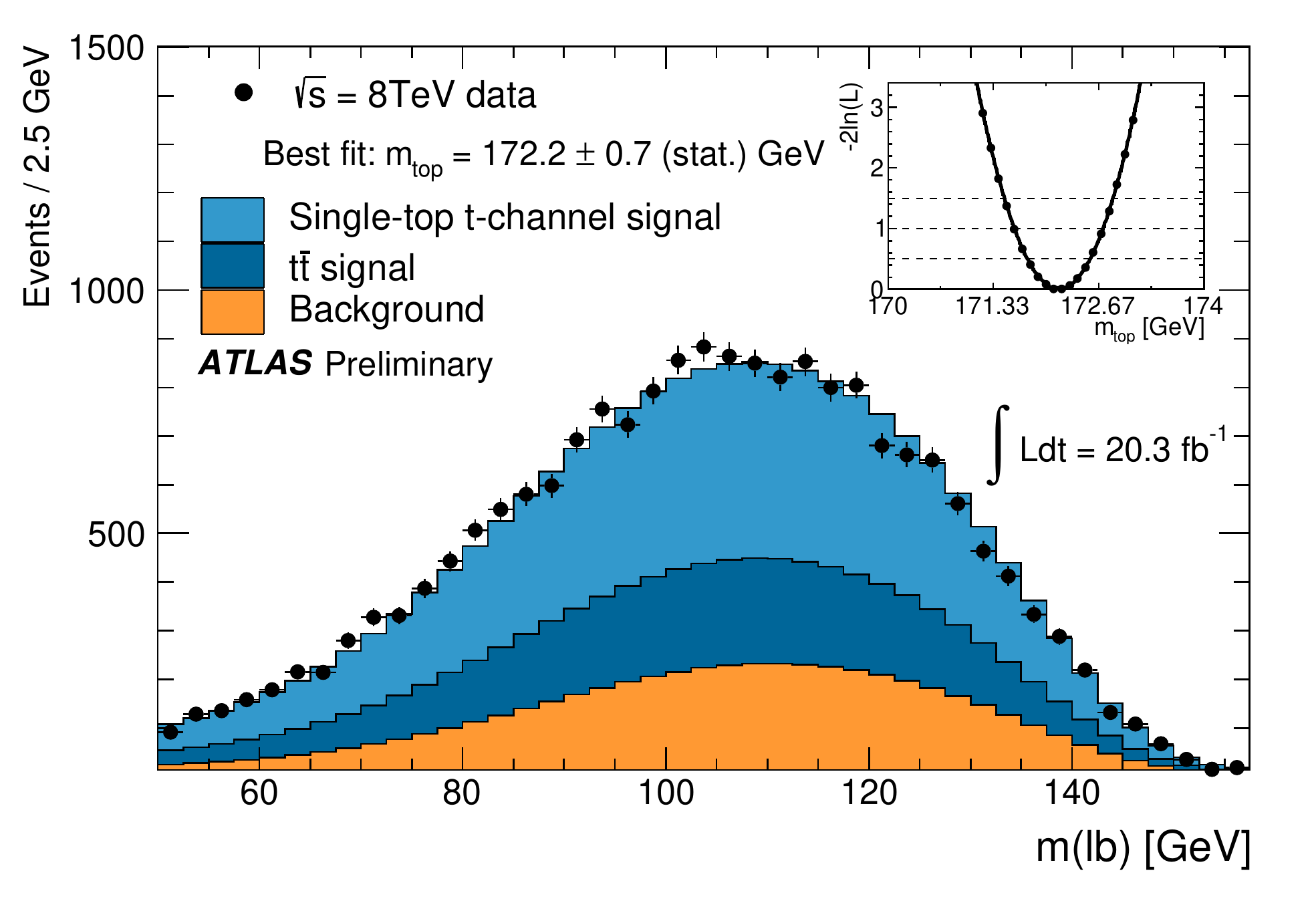}\hspace{2pc}%
\begin{minipage}[b]{20pc}\caption{\label{fig:datafit}Fitted \mlb{} distribution in data with the normalisation and \mtop{} being the best fit values. The relative mixture for the 
dominant single top $t$-channel production process and the other top processes, dominated by \ttbar{}, are shown in light and dark blue, respectively, and
correspond to the values determined in~\cite{ATLAS-CONF-2014-007}. The inset shows the corresponding $-2\ln \mathcal{L}$ profile as a function of the top quark mass~\cite{ATLAS-CONF-2014-055}.}
\end{minipage}
\end{figure}

The distribution of \mlb{} in the full dataset together with the corresponding fitted probability density functions for the signal and background is shown in Figure~\ref{fig:datafit}.\vspace*{-0.25cm}

The result has a total uncertainty of about $2\GeV$ which is dominated by systematic uncertainties. 
The largest contribution comes from JES uncertainties and the modelling of the $t$-channel process. 
Due to the $\ell + 2$-jet channel selection there is no statistical correlation between the 
dataset used in this analysis and any other analysis performed using the \ttbar{} final state.

The selection with exactly one tagged plus one untagged jet present in the final state leads to a reduced combinatorial 
background and better mass resolution compared to the $\ttbar \rightarrow $ lepton+jets or 
the \ttbar{} all hadronic decay channels. The presence of only one neutrino is an advantage with respect to 
the $\ttbar \rightarrow $ dilepton decay channel where the assignment of the missing transverse momentum 
to the neutrinos is ambiguous. 
These advantages in terms of systematics are complementary to the advantages of other channels,
e.g. the smaller contributions from backgrounds, indicating good prospects for combined measurements in the future.

\section*{Acknowledgements}

The author thankfully acknowledges the financial support of the BMBF (FSP101-ATLAS).

\smallskip

\end{document}